\documentclass[aps,preprint,showpacs,nofootinbib]{revtex4}
\usepackage{amsfonts}
\usepackage{amsmath}
\usepackage{amssymb}
\usepackage{relsize}
\usepackage{graphicx}
\usepackage{pdfpages}
\usepackage{subfigure}
\usepackage{float}
\usepackage[caption = false]{subfig}
\usepackage{color}

\usepackage{multirow}
\newcommand{\ic}{\textbf{\textbf{i}}}
\begin{document}

\title{Two-particle atomic coalescences: Boundary conditions for the Fock coefficient components}

\author{Evgeny Z. Liverts}
\affiliation{Racah Institute of Physics, The Hebrew University, Jerusalem 91904,
Israel}

\begin{abstract}
The exact values of the presently determined components
of the angular Fock coefficients
at the two-particle coalescences were obtained and systematized.
The Green Function approach was successfully applied to simplify the most complicated calculations.
The boundary conditions for the Fock coefficient components in the hyperspherical angular coordinates, which follows from
the Kato cusp conditions for the two-electron wave function in the natural interparticle coordinates, were derived.
The validity of the obtained boundary conditions was verified on examples of all the presently determined components.
The additional boundary conditions are not arising from the Kato cusp conditions were obtained as well.
The Wolfram Mathematica was used intensively.
\end{abstract}


\maketitle

\section{Introduction}\label{S0}

Investigation of the helium-like isoelectronic systems is a subject of a great number of physical articles.
However, only a few number of them (see, e.g., \cite{KATO,PBB,CDS,PNG,LEZM1,KNN,LEZ0}) were devoted to studying a specific states
of the two-particle coalescences, when the two electrons or one of the electrons and the nucleus fall into the same point
of the usual three-dimensional space.
An interest in such specific states is caused by the fact that they can serve as effective physical and mathematical models
for description of the various physical processes, such as, e.g., the two-electron photoionization \cite{LEZM2}.

The analytical treatment of the two-particle coalescences is based on
the Fock expansion \cite{FOCK} for the $^1S$ state wave function of the two-electron atom/ion
\begin{equation}\label{I1}
\chi(r_1,r_2,r_{12})\equiv\Psi(r,\alpha,\theta)=\sum_{k=0}^{\infty}r^k\sum_{p=0}^{[k/2]}\psi_{k,p}(\alpha,\theta)(\ln r)^p,
\end{equation}
where the hyperspherical angles $\alpha$ and $\theta$, and the  hyperspherical radius $r$ are expressed through the
interparticle distances $r_1,r_2$ and $r_{12}$ as follows
\begin{equation}\label{I2}
\alpha=2\arctan \left(r_2/r_1\right),~~~\theta=\arccos\left[(r_1^2+r_2^2-r_{12}^2)/(2r_1r_2)\right],
\end{equation}
\begin{equation}\label{I3}
r=\sqrt{r_1^2+r_2^2}.
\end{equation}
The works \cite{B37,MORG,LER,PLUV,FEM,AM1,GAM2,GM3,FOR} were devoted to investigation of the properties and to calculations of
the angular Fock coefficients (AFCs), $\psi_{k,p}(\alpha,\theta)$.
It has been proven that the AFCs satisfy (see, e.g., \cite{AM1} or \cite{LEZ1}) the Fock recurrence relation (FRR)
\begin{subequations}\label{I4}
\begin{align}
\left[ \Lambda^2-k(k+4)\right]\psi_{k,p}=h_{k,p},~~~~~~~~~~~~~~~~~~~~~~~~~~~~~~~~~~~~~~~~~~~~~~~~~~~~~~~~~~~~~\label{I4a}\\
h_{k,p}=2(k+2)(p+1)\psi_{k,p+1}+(p+1)(p+2)\psi_{k,p+2}-2 V \psi_{k-1,p}+2 E \psi_{k-2,p},\label{I4b}
\end{align}
\end{subequations}
where $E$ is the energy and $V\equiv V(\alpha,\theta)$ is the dimensionless Coulomb interaction for the two-electron atom/ion with infinitely massive nucleus of the charge $Z$.
The well-known hyperspherical harmonics $Y_{kl}(\alpha,\theta)$ (HH) are the eigenfunctions of the ($S$ state) hyperspherical angular momentum operator
$\Lambda^2$.

It was shown \cite{LEZ1} that any AFC, $\psi_{k,p}$ can be separated into the independent parts (components)
\begin{equation}\label{I11}
\psi_{k,p}(\alpha,\theta)=\sum_{j=p}^{k-p} \psi_{k,p}^{(j)}(\alpha,\theta) Z^j
\end{equation}
 associated with a definite power of $Z$, according to separation of the rhs (\ref{I4b})
of the FRR.
Accordingly, each of the FRRs (\ref{I4}) can be separated into the individual equations (IFRRs) for each component
\begin{equation}\label{I13}
\left[ \Lambda^2-k(k+4)\right]\psi_{k,p}^{(j)}(\alpha,\theta)=h_{k,p}^{(j)}(\alpha,\theta),
\end{equation}
where $h_{k,p}^{(j)}$ corresponds to the $Z$-power separation of the rhs (\ref{I4b}) which is similar to the separation formula (\ref{I11}).

\section{The exact values of the AFC components at the two-particle coalescences}\label{S1}

In the paper \cite{LEZ1} all of the AFC  components $\psi_{k,p}^{(j)}(\alpha,\theta)$ were presented for $k\leq 2$.
For $k>2$ the components $\psi_{3,0}^{(0)},~\psi_{3,0}^{(3)},~\psi_{3,1}^{(j_1)},~\psi_{4,1}^{(j_2)}$ and $\psi_{4,2}^{(j_3)}$
with all possible $j_1,j_2$ and $j_3$, and the subcomponents
$\psi_{3,0}^{(1a)},\psi_{3,0}^{(1b)},\psi_{3,0}^{(1c)},\psi_{3,0}^{(2a)},\psi_{3,0}^{(2b)},\psi_{3,0}^{(2c)}$
were presented, as well. The edge components $\psi_{k,0}^{(0)}$ and $\psi_{k,0}^{(k)}$ with $k\geq4$ as well as
subcomponent $\psi_{3,0}^{(2e)}$ were derived in \cite{LEZ2}.

In Tables \ref{T1} - \ref{T4} we present the exact values of the presently determined AFC components at the two-particle coalescences.
Notice, as we wrote "exact value" we meant representation expressed through the mathematical constants (like $\pi,\ln 2$ or the Catalan's
constant $G\simeq 0.915965...$) and rational numbers.
Remind that the \textbf{electron-nucleus} coalescence (ENC) line corresponds to the hyperspherical angles $\alpha=0\wedge\theta=\pi/2$,
whereas  $\alpha=\pi/2\wedge\theta=0$ describe the \textbf{electron-electron} coalescence (EEC).

There is no problem to perform the mentioned above calculations in case of the considered component was derived in form of the
explicit analytical expression.
However, in case of representation of the component/subcomponent by single series of the form
\begin{equation}\label{30}
\psi_{k,p}^{(j)}(\alpha,\theta)=\sum_{l=0}^\infty P_l(\cos\theta)(\sin \alpha)^l\omega_l^{(k,p;j)}(\alpha),
\end{equation}
(see, e.g., $\psi_{2,0}^{(1)}, \psi_{3,0}^{(2c)}, \psi_{3,0}^{(2e)}, \psi_{4,1}^{(2d)}$) the calculation of its exact value, at least at the EEC, is not a simple problem.
Therefore, we shall consider those calculations in details.

\subsection{Component $\psi_{2,0}^{(1)}$}\label{S1a}

There are three representations for the component $\psi_{2,0}^{(1p)}$ containing the different admixtures of the
$Y_{21}(\alpha,\theta)=\sin \alpha \cos \theta$. The first representation is the closed analytic expression presented
by Eq.(22)\cite{LEZ1}. Others are of the form
\begin{equation}\label{31}
\psi_{2,0}^{(1p)}=-\frac{1}{3}\left[\sin(\alpha/2)+\cos(\alpha/2)\right]\sqrt{1-\sin \alpha \cos\theta}+\chi_{20}(\alpha,\theta),
\end{equation}
where the function $\chi_{20}$ can be represented either by single series (second representation)
\begin{equation}\label{31a}
\chi_{20}(\alpha,\theta)=\sum_{l=0}^\infty P_l(\cos\theta)(\sin \alpha)^l\sigma_l(\rho)
\end{equation}
of the form (\ref{30}), or by a double series expansion in HHs.
The variable
\begin{equation}\label{31b}
\rho=\tan(\alpha/2)
\end{equation}
was introduced for convenience.
Only the first two cases are of the interest to us.
For the ENC, both representations give certainly the same results
\begin{equation}\label{32}
\lim_{\theta\rightarrow \frac{\pi}{2}}\lim_{\alpha\rightarrow 0}\psi_{2,0}^{(1)}(\alpha,\theta)=
\frac{1}{6}(\ln 2-3),
\end{equation}
because $Y_{21}(0,\pi/2)=0$.
One should emphasize that according to definition (\ref{30}) of the single series representation, the values of the corresponding component/subcomponent at the ENC ($\alpha=0$) equals $\omega_0^{(k,p;j)}(0)$.

At the EEC, one obtains
\begin{equation}\label{33}
\lim_{\theta\rightarrow 0}\lim_{\alpha\rightarrow \frac{\pi}{2}}\psi_{2,0}^{(1p)}(\alpha,\theta)=
\frac{1}{6}(1-\ln 2)
\end{equation}
for the analytic representation (22)\cite{LEZ1}. To obtain the "pure" solution (see \cite{LEZ1}), one needs to subtract the coefficient $\widetilde{C}_{21}^{(p)}\simeq 0.315837352$ from the result (\ref{33}),
taking into account that $Y_{21}\left(\frac{\pi}{2},0\right)=1$.
This admixture coefficient was computed by the expedient numerical integration with the integrand containing the complicated analytic expression (22) \cite{LEZ1}.
Then, making use of the expressions (63)-(65) from Ref.\cite{LEZ1} for $\sigma_l(\rho)$, one obtains for the subcomponent (\ref{31a}) at the EEC line:
\begin{equation}\label{34}
\chi_{20}\left(\frac{\pi}{2},0\right)=s_0+s_1,
\end{equation}
with
\begin{equation}\label{35}
s_0=\lim_{\rho\rightarrow 1}\left[\sigma_0(\rho)+ \sigma_1(\rho)\right]=
\frac{48G+\pi(3\pi-19-12 \ln 2)-14}{72 \pi},
\end{equation}
\begin{equation}\label{36}
s_1=\sum_{l=2}^{\infty}\sigma_l(1),
\end{equation}
where
\begin{equation}\label{37}
\sigma_l(1)=\frac{1}{2(l^2+l-2)}-\frac{(2l+1)\Gamma\left(\frac{l-1}{2}\right)\Gamma\left(\frac{l+1}{2}\right)}
{48\Gamma\left(\frac{l}{2}+1\right)\Gamma\left(\frac{l}{2}+2\right)},~~~~~(l\geq 2)
\end{equation}
and $G$ in the rhs of Eq.(\ref{35}) is the Catalan's constant.
Separating the summation (\ref{36}) by parts, one obtains:
\begin{equation}\label{38}
\sum_{l=2}^\infty\frac{1}{2(l^2+l-2)}=\frac{11}{36},
\end{equation}
\begin{equation}\label{39}
\sum_{l=2(2)}^\infty\frac{(2l+1)\Gamma\left(\frac{l-1}{2}\right)\Gamma\left(\frac{l+1}{2}\right)}
{48\Gamma\left(\frac{l}{2}+1\right)\Gamma\left(\frac{l}{2}+2\right)}=
\frac{\pi}{24},
\end{equation}
\begin{equation}\label{40}
\sum_{l=3(2)}^\infty\frac{(2l+1)\Gamma\left(\frac{l-1}{2}\right)\Gamma\left(\frac{l+1}{2}\right)}
{48\Gamma\left(\frac{l}{2}+1\right)\Gamma\left(\frac{l}{2}+2\right)}=
\frac{2}{9\pi}.
\end{equation}
Combining the results (\ref{34})-(\ref{40}), one obtains
\begin{equation}\label{41}
\chi_{20}\left(\frac{\pi}{2},0\right)=\frac{16 G +\pi(1-4\ln 2)-10}{24\pi}.
\end{equation}
 The "pure" component can be obtained by subtracting the admixture coefficient $C_{21}^{(p)}=(\pi+4)/9\pi$ (see the end Sec.VI \cite{LEZ1}) from the rhs of Eq.(\ref{41}).
Thus, the final result for the EEC is
\begin{equation}\label{42}
\psi_{2,0}^{(1)}\left(\frac{\pi}{2},0\right)=\frac{48 G -62+\pi(5+12\ln 2)}{72\pi}.
\end{equation}
Comparing the exact EEC value of the above component derived with using the single series and the closed analytic representations, one obtains the exact value of the admixture coefficient
\begin{equation}\label{43}
\widetilde{C}_{21}^{(p)}=\frac{62+17\pi-48G}{72\pi}\simeq 0.315837352.
\end{equation}

\subsection{Component $\psi_{4,1}^{(2)}$}\label{S1b}

The component $\psi_{4,1}^{(2)}$ was separated \cite{LEZ1} into the parts
\begin{equation}\label{44a}
\psi_{4,1}^{(2)}=\psi_{4,1}^{(2b)}+\psi_{4,1}^{(2c)}+\psi_{4,1}^{(2d)},
\end{equation}
where  $\psi_{4,1}^{(2b)}$ is presented in Table 1 of Ref.\cite{LEZ1}, and $\psi_{4,1}^{(2c)}$ is defined by Eq.(92)\cite{LEZ1}.
For these subcomponents calculated at the two-particle coalescences, one easily obtains:
\begin{equation}\label{44}
\psi_{4,1}^{(2b)}\left(0,\frac{\pi}{2}\right)=\frac{(\pi-2)(5\pi-14)}{1080\pi^2},~~~
\psi_{4,1}^{(2b)}\left(\frac{\pi}{2},0\right)=\frac{(\pi-2)(5\pi-14)}{3240\pi^2},~~~
\end{equation}
\begin{equation}\label{45}
\psi_{4,1}^{(2c)}\left(0,\frac{\pi}{2}\right)=0,~~~
\psi_{4,1}^{(2c)}\left(\frac{\pi}{2},0\right)=-\frac{37(\pi-2)(5\pi-14)}{8100\pi^2}.~~~
\end{equation}
The subcomponent $\psi_{4,1}^{(2d)}$ was represented by single series of the form (\ref{30}) with function $\omega_l^{(4,1;2d)}(\alpha)\equiv \tau_l(\rho)$ defined by Eqs.(94)-(96)\cite{LEZ1} for $l=0,1,2$. The explicit analytic expression for $\tau_l$ with $l\geq3$ was represented in \cite{LEZ2}. According to representation (\ref{30}) and Eq.(94)\cite{LEZ1}, one obtains
\begin{equation}\label{46}
\psi_{4,1}^{(2d)}\left(0,\frac{\pi}{2}\right)=\tau_0(0)=
\frac{(\pi-2)\left[247-300 G+50\pi(3\ln2-2)\right]}{2700\pi^2}
\end{equation}
 at the ENC, and
\begin{equation}\label{47}
\psi_{4,1}^{(2d)}\left(\frac{\pi}{2},0\right)=\sum_{l=0}^\infty\tau_l(1)
\end{equation}
at the EEC. The explicit representations (see Sec.VII \cite{LEZ1}) for $l=0,1,2$ yields:
\begin{equation}\label{48}
\tau_0(1)=\frac{(2-\pi)\left[247-300 G+5\pi(15\ln2-16)\right]}{8100\pi^2},
\end{equation}
\begin{equation}\label{49}
\tau_1(1)=\frac{(\pi-2)(735\pi-5788)}{50400\pi},
\end{equation}
\begin{equation}\label{50}
\tau_2(1)=\frac{(\pi-2)\left[4592-16800 G+5\pi(109+840\ln2)\right]}{113400\pi^2}.
\end{equation}
The most effective way to calculate $\tau_l(1)$ with $(l>2)$ is use of the formulas \cite{LEZ1}
\begin{equation}\label{51}
\tau_l(1)=\frac{u_{4l}(1)}{2l+1}\int_0^1 \frac{v_{4l}(t)h_l(t)t^{2l+2}}{(1+t^2)^{2l+3}}dt+A_2(l)v_{4l}(1),
\end{equation}
where
\begin{subequations}\label{52}
\begin{align}
u_{4l}(\rho)=\rho^{-2l-1}(\rho^2+1)^{l+4}~_2F_1\left(\frac{7}{2},3-l;\frac{1}{2}-l;-\rho^2\right),~~~~~~~\label{6a}\\
v_{4l}(\rho)=(\rho^2+1)^{l+4}~_2F_1\left(\frac{7}{2},4+l;l+\frac{3}{2};-\rho^2\right),~~~~~~~~~~~~~~~~\label{6b}
\end{align}
\end{subequations}
\begin{eqnarray}\label{53}
h_l(\rho)=-\frac{(\pi-2)(\rho+1)\left(\rho^2+1\right)^{l-1}}{3\pi(2l-1)(2l+3)2^{l+1}}\times
\nonumber~~~~~~~~~~~~~~~~~~~~~~~~~~~~~~~~~~~~~~~~~~~~~~~~~\\
\left[\frac{15-4l(l+1)(4l+11)}{(2l-3)(2l+5)\rho}+4l(2l+3)+
2\rho+4(l+1)(2l-1)\rho^2+\frac{(2l-1)(4l+5)\rho^3}{2l+5}\right].~~~
\end{eqnarray}
The coefficient $A_2(l)$ equals zero for odd $l$, whereas for even $l$ (see \cite{LEZ2})
\begin{eqnarray}\label{54}
A_2(l)=\frac{(2-\pi)}{360 l (l-2)\Gamma(l+\frac{1}{2})\pi^{3/2}}\times
~~~~~~~~~~~~~~~~~~~~~~~~~~~~~~~~~~~~~~~~~~~~~~~~~~~~~~~~~~~~~~~~\nonumber~\\
\left\{
\frac{\left[l(l+1)(688 l^4+1376 l^3-2480 l^2-3168 l+465)+450\right]\Gamma\left(\frac{l-1}{2}\right)\Gamma\left(\frac{l+1}{2}\right)}
{(2l-3)(2l-1)(2l+1)(2l+3)(2l+5)}-\frac{56}{l-1}\left(\frac{l}{2}!\right)^2
\right\}.~~~
\end{eqnarray}
The most simple method of finding a simple representation for $\tau_l(1)$ with $l\geq3$ involves the use of the \emph{Mathematica} operator
 \textbf{FindSequenceFunction} (see examples in \cite{LEZ2}).
In particular, using Eqs.(\ref{51})-(\ref{54}) we have calculated $\tau_l(1)$ for $2<l<30$. It was found that
\begin{equation}\label{55}
\tau_l(1)=\frac{\pi-2}{\pi}
\left\{ \begin{array}{c}~~
\mathlarger{(a_l+\pi b_l)},~~~~~~~~~~~~~for~odd~l\\
\mathlarger{(\widetilde{a}_l+\pi \widetilde{b}_l)/\pi},~~~~~~~~~~for~even~l
\end{array}\right.
\end{equation}
where $a_l,b_l,\widetilde{a}_l,\widetilde{b}_l$ are rational numbers.
Making use of the calculated sequences for each of the coefficients $a_l,b_l,\widetilde{a}_l$ and $\widetilde{b}_l$,
the \emph{Mathematica} operator \textbf{FindSequenceFunction} enables us to find the general forms of these coefficients as functions of $l$.
Notice that for a given sequence there is a minimal number of terms to enable \emph{Mathematica} to find the formula of the general term.
In particular, for the coefficients $a_l$ and $\widetilde{b}_l$ these minimal number is 16, whereas for $\widetilde{a}_l$ and $b_l$ it equals 10.
Thus, finally one obtains:
\begin{eqnarray}\label{56}
\tau_l(1)=\frac{2-\pi}{180\pi(l+1)(l+3)}\times
~~~~~~~~~~~~~~~~~~~~~~~~~~~~~~~~~~~~~~~~~~~~~~~~~~~~~~~~~~~~~~~~~~~~~~~~~~~~~~~~\nonumber~\\
\left\{\frac{l(l+1)[16l(l+1)(43l(l+1)-198)+465]+450}{(l-2)l(2l-3)(2l-1)(2l+3)(2l+5)}-
\frac{7(2l+1)\Gamma\left(\frac{l}{2}-1\right)\Gamma\left(\frac{l}{2}+1\right)}{\Gamma^2\left(\frac{l+1}{2}\right)}
\right\}.~~~~~~~
\end{eqnarray}
The expedient summation by \emph{Mathematica} gives
\begin{equation}\label{57}
\sum_{l=3}^\infty \frac{l(l+1)[16l(l+1)(43l(l+1)-198)+465]+450}{(l-2)l(l+1)(l+3)(2l-3)(2l-1)(2l+3)(2l+5)}=\frac{675}{35},
\end{equation}
\begin{equation}\label{58}
\sum_{l=3(2)}^\infty\frac{(2l+1)\Gamma\left(\frac{l}{2}-1\right)\Gamma\left(\frac{l}{2}+1\right)}
{(l+1)(l+3)\Gamma^2\left(\frac{l+1}{2}\right)} =\frac{3\pi}{8},
\end{equation}
\begin{equation}\label{59}
\sum_{l=4(2)}^\infty\frac{(2l+1)\Gamma\left(\frac{l}{2}-1\right)\Gamma\left(\frac{l}{2}+1\right)}
{(l+1)(l+3)\Gamma^2\left(\frac{l+1}{2}\right)} =\frac{32}{15\pi}.
\end{equation}
Notice that figure $2$ in brackets following the lower limit of summation denotes its step (by default a step equals $1$).
Summarising the results (\ref{47})-(\ref{59}), one obtains
\begin{equation}\label{60}
\psi_{4,1}^{(2d)}\left(\frac{\pi}{2},0\right)=
\frac{\pi-2}{75600\pi^2}\left[
7028-8400 G+3\pi(735\pi-5228+700\ln 2)
\right].
\end{equation}
Finally, summation (\ref{44a}) of the proper subcomponents yields
\begin{equation}\label{61}
\psi_{4,1}^{(2)}\left(0,\frac{\pi}{2}\right)=
\frac{(\pi-2)\left[424-600 G+25\pi(12\ln2-7)\right]}{5400\pi^2}
\end{equation}
 at the ENC, and
\begin{equation}\label{62}
\psi_{4,1}^{(2)}\left(\frac{\pi}{2},0\right)=
\frac{\pi-2}{75600\pi^2}\left[
11536-8400 G+\pi(2205\pi-17294+2100\ln 2)
\right].
\end{equation}
 at the EEC.

\subsection{Subcomponent $\psi_{3,0}^{(2c)}$}\label{S1c}

The subcomponent $\psi_{3,0}^{(2c)}$ was represented by single series of the form (\ref{30}) with function
$\omega_l^{(3,0;2c)}(\alpha)\equiv \phi_l(\rho)$ defined \cite{LEZ1} as follows
\begin{equation}\label{64}
\phi_l(\rho)=\phi_l^{(p)}(\rho)+c_l v_{3l}(\rho).
\end{equation}
The functions included into the rhs of Eq.(\ref{64}) are
\begin{equation}\label{65}
v_{3l}(\rho)=\left(\rho^2+1\right)^{l-\frac{3}{2}}
\left[\frac{(2l-3)(2l-1)}{(2l+3)(2l+5)}\rho^4+\frac{2(2l-3)}{2l+3}\rho^2+1\right],
\end{equation}
\begin{equation}\label{66}
\phi_l^{(p)}(\rho)=\frac{2^{-l}\left(\rho^2+1\right)^{l-\frac{3}{2}}}{3(2l-3)(2l-1)(2l+3)(2l+5)}
\left[2f_{1l}(\rho)+\frac{2f_{2l}(\rho)+f_{3l}(\rho)}{2l+1}\right],
\end{equation}
where
\begin{equation}\label{67}
f_{1l}(\rho)=\left[9-4l(l+2)\right]\rho+\left(13-4l^2\right)\rho^3,~~~~~~~~~~~~~~~~~~~~~~~~~~~~~~~~~~~~
\end{equation}
\begin{equation}\label{68}
f_{2l}(\rho)=\left[(2l-3)(2l-1)\rho^4+2(2l-3)(2l+5)\rho^2+(2l+3)(2l+5)\right]\arctan(\rho),~~~~~~~~~
\end{equation}
\begin{eqnarray}\label{69}
f_{3l}(\rho)=-\left[(2l+3)(2l+5)\rho^4+2(2l-3)(2l+5)\rho^2+(2l-3)(2l-1)\right]\times
\nonumber~~~~~~~~~~~~~~~~~~~~\\
\frac{\rho}{l+1}~_2F_1\left(1,l+1;l+2;-\rho^2\right).~~~~~~~~~~~~~~~~~
\end{eqnarray}
Simple representation for the coefficient $c_l$ was derived in Ref.\cite{LEZ2} in the form
\begin{equation}\label{70}
c_l=\frac{2(2l+1)-\pi-H_{\frac{l}{2}}+H_{\frac{l-1}{2}}}{6(2l-3)(2l-1)(2l+1)2^l},
\end{equation}
where $H_z$  is the harmonic number.

According to the single series representation (\ref{30}), and making use of Eqs.(\ref{64})-(\ref{70}), one obtains:
\begin{equation}\label{72}
\psi_{3,0}^{(2c)}\left(0,\frac{\pi}{2}\right)=\phi_0(0)=
\frac{1}{18}(2-\pi-2\ln 2)
\end{equation}
at the ENC, and
\begin{equation}\label{73}
\psi_{3,0}^{(2c)}\left(\frac{\pi}{2},0\right)=\sum_{l=0}^\infty\phi_l(1)
\end{equation}
at the EEC. To obtain a simple representation for $\phi_l(1)$, we use again the \emph{Mathematica} operator \textbf{FindSequenceFunction} (see the previous Subsection and \cite{LEZ2}). The  \emph{Mathematica} calculations of $\phi_l(1)$ with given $l\geq0$, performed on the basis of Eqs.(\ref{64})-(\ref{70}), show that
\begin{equation}\label{74}
\phi_l(1)=\sqrt{2}(\textmd{a}_l+\textmd{b}_l\ln 2),
\end{equation}
where $\textmd{a}_l$ and $\textmd{b}_l$ are rational numbers.
It is enough to use the sequence of  $\textmd{a}_l$ with $0\leq l\leq 27$, and the sequence of  $\textmd{b}_l$ with $0\leq l\leq 6$ to find
\begin{equation}\label{75}
\textmd{a}_l=\frac{1}{3(2l-3)(2l+1)(2l+5)}
\left[\frac{8(2l+1)}{4l(l+1)-3}+H_\frac{l-1}{2}-H_\frac{l}{2}+(-1)^l2\ln2
\right],
\end{equation}
\begin{equation}\label{76}
\textmd{b}_l=-\frac{2(-1)^l}{3(2l-3)(2l+1)(2l+5)}.
\end{equation}
Whence, one easily obtains
\begin{equation}\label{77}
\phi_l(1)=\frac{\sqrt{2}}{3(2l-3)(2l+1)(2l+5)}\left[\frac{8(2l+1)}{4l(l+1)-3}+H_\frac{l-1}{2}-H_\frac{l}{2}\right].
\end{equation}
\emph{Mathematica} summations yield
\begin{equation}\label{78}
\sum_{l=0}^\infty\frac{1}{(2l-3)(2l+5)[4l(l+1)-3]}=0,
\end{equation}
\begin{equation}\label{79}
\sum_{l=0}^\infty\frac{1}{(2l-3)(2l+1)(2l+5)}\left(H_\frac{l-1}{2}-H_\frac{l}{2}\right)=\frac{1}{9}.
\end{equation}
Thus, finally for the EEC, one obtains:
\begin{equation}\label{73}
\psi_{3,0}^{(2c)}\left(\frac{\pi}{2},0\right)=\frac{\sqrt{2}}{27}.
\end{equation}

\subsection{Subcomponent $\psi_{3,0}^{(2e)}$}\label{S1d}

In Ref.\cite{LEZ2} the subcomponent $\psi_{3,0}^{(2e)}$ was derived in the form of the single series (\ref{30}) with function
$\omega_l^{(3,0;2e)}(\alpha)\equiv \lambda_l(\rho)$ defined as
\begin{equation}\label{11}
\lambda_l(\rho)=\frac{1}{2l+1}\left\{u_{3l}(\rho)\mathcal{V}_{3l}(\rho)-v_{3l}(\rho)
\left[\mathcal{U}_{3l}(\rho)-(2l+1)s_l\right]\right\},
\end{equation}
where
\begin{equation}\label{11a}
s_l=\frac{2^{-l-3}}{(2l-3)(2l-1)(2l+1)}\left[2l(l+1)\left(H_{\frac{l+1}{2}}-H_{\frac{l}{2}}-\pi\right)+2l+3\right\},
\end{equation}
\begin{equation}\label{12}
u_{3l}(\rho)=\frac{\left(\rho^2+1\right)^{l-\frac{3}{2}}}{\rho^{2l+1}}
\left[\frac{(2l+3)(2l+5)}{(2l-3)(2l-1)}\rho^4+\frac{2(2l+5)}{2l-1}\rho^2+1\right],
\end{equation}
\begin{equation}\label{13}
\mathcal{U}_{3l}(\rho)=
-\frac{l(l+1)\left[(\rho^2+1)^4\arctan (\rho)+\rho^7-\rho\right]+(l^2-7l-10)\rho^5-(l^2+9l-2)\rho^3}
{2^l(2l-3)(2l-1)(\rho^2+1)^4},
\end{equation}
\begin{eqnarray}\label{14}
\mathcal{V}_{3l}(\rho)=
~\nonumber~~~~~~~~~~~~~~~~~~~~~~~~~~~~~~~~~~~~~~~~~~~~~~~~~~~~~~~~~~~~~~~~~~~~~~~~~~~~~~~~~~~~~~~~~~~~~~~~~~~~~~~\\
-\left[(-2)^l(l-2)(l-1)(2l+3)(2l+5)\right]^{-1}
\left\{12\left[B_{-\rho^2}(l+1,-3)-B_{-\rho^2}(l+1,-4)\right]+
\right.
~~~\nonumber~\\
\left.
(2l-3)\rho^2\left[2l^2+l-7+(l-2)(2l-1)\rho^2\right]\left[(3-l)B_{-\rho^2}(l+1,-3)-4B_{-\rho^2}(l+1,-4)\right]
\right\}
.~~~~
\end{eqnarray}
Here $B_z(a,b)$ is the Euler beta function.
It is seen that expression (\ref{14}) cannot be applied directly for $l=1,2$. For this values of $l$, we have
\begin{equation}\label{15}
\mathcal{V}_{31}(\rho)=
\frac{1}{140}\left[\frac{3+10\rho^2+11\rho^4-20\rho^6}{(1+\rho^2)^4}+2\ln(1+\rho^2)\right],~~~~~~~~~~
\end{equation}
\begin{equation}\label{16}
\mathcal{V}_{32}(\rho)=
-\frac{1}{84}\left[\frac{5+14\rho^2+9\rho^4-6\rho^6+24\rho^8+6\rho^{10}}{6(1+\rho^2)^4}+\ln(1+\rho^2)\right].
\end{equation}
Function $v_{3l}(\rho)$ is defined by Eq.(\ref{65}).

For the ENC one obtains
\begin{equation}\label{17}
\psi_{3,0}^{(2e)}\left(0,\frac{\pi}{2}\right)=\lambda_0(0)=\frac{1}{8},
\end{equation}
whereas for the ENC, the single-series representation (\ref{30}) yields
\begin{equation}\label{18}
\psi_{3,0}^{(2e)}\left(\frac{\pi}{2},0\right)=\sum_{l=0}^\infty\lambda_l(1).
\end{equation}
Using formulas (\ref{11})-(\ref{16}), one obtains
\begin{equation}\label{19}
\lambda_l(1)=\frac{2l(l+1)\left(H_{\frac{l+1}{2}}-H_{\frac{l}{2}}\right)-6l-1}
{2\sqrt{2}(2l-3)(2l+1)(2l+5)}
\end{equation}
Then using the integral representation for the harmonic number
\begin{equation*}
H_z=\int_0^1 \frac{1-t^z}{1-t}dt,~~~~~~~~~~Re(z)>-1
\end{equation*}
and making use of the change of variable $x=\sqrt{t}$, one obtains
\begin{equation}\label{20}
\sum_{l=0}^\infty\lambda_l(1)=\frac{1}{2\sqrt{2}}\left(4S_1-S_2\right),
\end{equation}
where
\begin{equation}\label{21}
S_2=\sum_{l=0}^\infty \frac{6l+1}{(2l-3)(2l+1)(2l+5)}=\frac{1}{6},
\end{equation}
\begin{eqnarray}\label{21a}
S_1=\frac{1}{2}\sum_{l=0}^\infty \frac{l(l+1)\left(H_{\frac{l+1}{2}}-H_{\frac{l}{2}}\right)}{(2l-3)(2l+1)(2l+5)}=
\int_0^1\left(\sum_{l=0}^\infty\frac{l(l+1)x^{l+1}}{(2l-3)(2l+1)(2l+5)}\right)\frac{dx}{x+1}=
~~~\nonumber~\\
\int_0^1\frac{\text{arctanh} (\sqrt{x})(15x^4+2x^2+15)-5\sqrt{x}(3x^3+x^2+x+3)}
{128 x^{3/2}}\left(\frac{dx}{x+1}\right)=\frac{1-G}{4}.~~~~~~~~~~~
\end{eqnarray}
$G$ as previously denotes the Catalan's constant. Inserting the results (\ref{20})-(\ref{21a}) into the rhs of Eq.(\ref{18}), one finally obtains
\begin{equation}\label{21b}
\psi_{3,0}^{(2e)}\left(\frac{\pi}{2},0\right)=\frac{5-6 G}{12\sqrt{2}}.
\end{equation}

\section{Green functions approach}\label{S2}

The Green functions approach to calculation of the AFCs was presented in \cite{ABT}.
It follows from the general formulas that any component/subcomponent of the AFC at the electron-nucleus coalescence can be calculated by the following simple way:
\begin{equation}\label{201}
\psi_{k,p}^{(j)}\left(0,\frac{\pi}{2}\right)=\frac{1}{8}\int_0^{\pi}d \alpha \sin \alpha
\cos\left[\left(\frac{k}{2}+1\right)\alpha\right]\zeta(\alpha)
\int_0^{\pi}d \theta \sin \theta h_{k,p}^{(j)}(\alpha,\theta),
\end{equation}
where
\begin{equation}\label{202}
\zeta(\alpha)=
\left\{ \begin{array}{c}
\mathlarger{1},~~~~~~~~~~~~~k~~odd\\
\mathlarger{1-\frac{\alpha}{\pi}}.~~~~~~~~k~~even\\
\end{array}\right.
\end{equation}
For the electron-electron coalescence, one obtains the more complicated formula of the form
\begin{equation}\label{203}
\psi_{k,p}^{(j)}\left(\frac{\pi}{2},0\right)=\frac{1}{8}\int_0^{\pi}d \alpha \sin^2 \alpha
\int_0^{\pi}d \theta \sin \theta h_{k,p}^{(j)}(\alpha,\theta)\cos\left[\left(\frac{k}{2}+1\right)\gamma\right]\frac{\zeta(\gamma)}{\sin \gamma},
\end{equation}
where the angle $\gamma$ is defined by the relation
\begin{equation}\label{204}
\cos\gamma=\sin \alpha \cos \theta.
\end{equation}
It is clear that $h_{k,p}^{(j)}$ in Eqs.(\ref{201}),(\ref{203}) represents the rhs of the corresponding IFRR (\ref{I13}).
It is seen that Eq.(\ref{201}) can be obtained by changing $\gamma$ for $\alpha$ in Eq.(\ref{203}).
Notice that the above formulas with even $k$ are correct only for the so called "pure" components/subcomponents (see \cite{LEZ1}).
Using  Eqs.(\ref{201})-(\ref{204}) we have recalculated all of the AFC (presented in \cite{LEZ1,LEZ2}) at the two-particle coalescences.
The results were coincident with direct derivations based on the explicit representations (including single-series ones) of the AFC.

The component $\psi_{3,0}^{(2)}(\alpha,\theta)$ was not obtained previously, because of the difficulties with calculations of the subcomponent $\psi_{3,0}^{(2d)}$
which represents the physical solution of the IFRR
\begin{equation}\label{205}
\left(\Lambda^2-21\right)\psi_{3,0}^{(2d)}(\alpha,\theta)=h_{3,0}^{(2d)}(\alpha,\theta),
\end{equation}
with the rhs
\begin{equation}\label{206}
h_{3,0}^{(2d)}(\alpha,\theta)=\frac{4}{\sin \alpha}\left[\sin\left(\frac{\alpha}{2}\right)+\cos\left(\frac{\alpha}{2}\right)\right]
\chi_{20}(\alpha,\theta),
\end{equation}
where the function $\chi_{20}$ is defined by the single series (\ref{31a}). The use of Eq.(\ref{201}) enables us to obtain
the exact value of the mentioned subcomponent at the ENC as follows
\begin{eqnarray}\label{207}
\psi_{3,0}^{(2d)}\left(0,\frac{\pi}{2}\right)
=\frac{1}{8}\int_0^{\pi}d \alpha \sin \alpha
\cos\left(\frac{5\alpha}{2}\right)
\int_0^{\pi}d \theta \sin \theta h_{3,0}^{(2d)}(\alpha,\theta)=
~~~~~~~~~~~~~~~~~~~~~~~~~~~~\nonumber~\\
\int_0^{\pi/2}\left[\sin\left(\frac{\alpha}{2}\right)+\cos\left(\frac{\alpha}{2}\right)\right]
\cos\left(\frac{5\alpha}{2}\right)\sigma_0(\alpha)d\alpha+
~~~~~~~~~~~~~~~~~~~\nonumber~\\
\int_{\pi/2}^\pi\left[\sin\left(\frac{\alpha}{2}\right)+\cos\left(\frac{\alpha}{2}\right)\right]
\cos\left(\frac{5\alpha}{2}\right)\sigma_0(\pi-\alpha)d\alpha=
\frac{1}{288}\left[24-48 G+\pi(16-3\pi)\right],~~
\end{eqnarray}
where (see \cite{LEZ1})
\begin{equation}\label{208}
\sigma_0(\alpha)=\frac{1}{12}\left\{\left(2\sin \alpha-\frac{1}{\sin \alpha}\right)\alpha+
\cos \alpha\left[2\ln(\cos \alpha+1)+1\right]-\sin \alpha-2
\right\}.~~~~0\leq\alpha\leq\pi/2
\end{equation}
To use $\sigma_0(\alpha)$ in the range $\pi/2<\alpha\leq\pi$, one should replace $\alpha$ by $\pi-\alpha$ in the rhs of Eq.(\ref{208}).

On the other hand, the exact value of subcomponent $\psi_{3,0}^{(2d)}$ at the ENC can be derived by use of the single-series representation
\begin{equation}\label{209}
\psi_{3,0}^{(2d)}(\alpha,\theta)=\sum_{l=0}^\infty P_l(\cos\theta)(\sin \alpha)^lg_l(\rho),
\end{equation}
which gives
\begin{equation}\label{210}
\psi_{3,0}^{(2d)}(0,\frac{\pi}{2})=g_0(0).
\end{equation}
We have obtained the exact analytic representation (see Appendix \ref{SA}) for the function $g_0(\rho)$
which at the ENC $(\rho\rightarrow 0)$ is certainly coincident with the result (\ref{207}).

Now we can calculate the exact value of the component $\psi_{3,0}^{(2)}$ at the ENC. Gathering the subcomponents, one obtains
\begin{equation}\label{211}
\left.\psi_{3,0}^{(2)}\left(0,\frac{\pi}{2}\right)=
\left(\psi_{3,0}^{(2a)}+\psi_{3,0}^{(2b)}+\psi_{3,0}^{(2c)}+\psi_{3,0}^{(2d)}+\psi_{3,0}^{(2e)}\right)\right|_{\alpha=0,\theta=\frac{\pi}{2}}=
\frac{124-48G-3\pi^2-32\ln 2}{288}.
\end{equation}
All the subcomponents along with the resulting component (\ref{211}) at the ENC are presented in Table \ref{T4}.
The exact calculation of the component  $\psi_{3,0}^{(2)}$ at the EEC is still difficult.

\section{the boundary conditions for the AFC components}\label{S3}

The Kato cusp condition (KCC) \cite{KATO} for the two-electron atomic wave function (\ref{I1}) at the ENC reads
\begin{equation}\label{81}
\left.\frac{\partial \chi(r_1,r_2,r_{12})}{\partial r_2}\right|_{r_2=0}=-Z \chi(r,0,r),
\end{equation}
where according to definition (\ref{I3}), $r=r_1=r_{12}$ is the hyperspherical radius (\ref{I3}) at the ENC line.
The trivial chain rule relation yields
\begin{equation}\label{82}
\frac{\partial \Psi(r,\alpha,\theta)}{\partial r_2}=
\frac{\partial\Psi}{\partial r}\frac{\partial r}{\partial r_2}+
\frac{\partial\Psi}{\partial \alpha}\frac{\partial \alpha}{\partial r_2}+
\frac{\partial\Psi}{\partial \theta}\frac{\partial \theta}{\partial r_2},
\end{equation}
where according to Eqs.(\ref{I2}), (\ref{I3}), one obtains for the ENC
\begin{subequations}\label{83}
\begin{align}
\left.\frac{\partial r}{\partial r_2}\right|_{r_2=0}=0,~~~~~~~~~~~~~~~~~~~~~~~~~\label{83a}\\
\left.\frac{\partial \alpha}{\partial r_2}\right|_{r_2=0}=\frac{2}{r},~~~~~~~~~~~~~~~~~~~~~~~~~~\label{83b}\\
\left.\frac{\partial \theta}{\partial r_2}\right|_{r_2=0}=-\frac{1}{2r}.~~~~~~~~~~~~~~~~~~~~~~~~~~\label{83b}
\end{align}
\end{subequations}
Inserting the the Fock expansion (\ref{I1}) into the KCC (\ref{81}), turning to the hyperspherical coordinates by Eqs.(\ref{82})-(\ref{83}),
and using the $Z$-power separation (\ref{I11}), one obtains
\begin{eqnarray}\label{84}
\sum_{k=0}^\infty r^{k-1}\sum_{p=0}^{[k/2]}\ln^p r \sum_{j=p}^{k-p}Z^j
\left.\left(2\frac{\partial\psi_{k,p}^{(j)}(\alpha,\theta)}{\partial \alpha}
-\frac{1}{2}\frac{\partial\psi_{k,p}^{(j)}(\alpha,\theta)}{\partial \theta}
\right)\right|_{\alpha=0,\theta=\pi/2}=
~~~~~~~~~~~~~~~~~~~~~\nonumber~\\
-\sum_{k=0}^\infty r^{k}\sum_{p=0}^{[k/2]}\ln^p r \sum_{j=p}^{k-p}Z^{j+1}
\psi_{k,p}^{(j)}\left(0,\frac{\pi}{2}\right).~~~~~~~
\end{eqnarray}
Equating coefficients for the same powers of $r,\ln r$ and $Z$ in both sides of Eq.(\ref{84}), one obtains the following equation
\begin{equation}\label{85}
\psi_{k,p}^{(j)}\left(0,\frac{\pi}{2}\right)=\frac{1}{2}
\left.\frac{\partial\psi_{k+1,p}^{(j+1)}(\alpha,\theta)}{\partial \theta}\right|_{\alpha=0,\theta=\pi/2}-
2\left.\frac{\partial\psi_{k+1,p}^{(j+1)}(\alpha,\theta)}{\partial \alpha}\right|_{\alpha=0,\theta=\pi/2}
\end{equation}
for the AFC components at the ENC line.

In its turn, the KCC for the two-electron atomic wave function (\ref{I1}) at the EEC is
\begin{equation}\label{86}
\left.\frac{\partial \chi(r_1,r_2,r_{12})}{\partial r_{12}}\right|_{r_{12}=0}=
\frac{1}{2} \chi\left(\frac{r}{\sqrt{2}},\frac{r}{\sqrt{2}},0\right),
\end{equation}
where $r_1=r_2=r/\sqrt{2}$.
Turning to the hyperspherical coordinates, one obtains
\begin{equation}\label{87}
\frac{\partial \Psi(r,\alpha,\theta)}{\partial r_{12}}=
\frac{\partial\Psi}{\partial r}\frac{\partial r}{\partial r_{12}}+
\frac{\partial\Psi}{\partial \alpha}\frac{\partial \alpha}{\partial r_{12}}+
\frac{\partial\Psi}{\partial \theta}\frac{\partial \theta}{\partial r_{12}},
\end{equation}
where according to Eqs.(\ref{I2}), (\ref{I3}), we have for the EEC line
\begin{subequations}\label{88}
\begin{align}
\left.\frac{\partial r}{\partial r_{12}}\right|_{r_{12}=0}=0,~~~~~~~~~~~~~~~~~~~~~~~~~\label{83a}\\
\left.\frac{\partial \alpha}{\partial r_{12}}\right|_{r_{12}=0}=0,~~~~~~~~~~~~~~~~~~~~~~~~~~\label{83b}\\
\left.\frac{\partial \theta}{\partial r_{12}}\right|_{r_{12}=0}=\frac{\sqrt{2}}{r}.~~~~~~~~~~~~~~~~~~~~~~~~\label{83b}
\end{align}
\end{subequations}
Inserting the the Fock expansion (\ref{I1}) into the KCC (\ref{86}), turning to the hyperspherical coordinates by Eqs.(\ref{87})-(\ref{88}), and using the $Z$-power separation (\ref{I11}), one obtains
\begin{equation}\label{89}
\sqrt{2}\sum_{k=0}^\infty r^{k-1}\sum_{p=0}^{[k/2]}\ln^p r \sum_{j=p}^{k-p}Z^j
\left.\frac{\partial\psi_{k,p}^{(j)}(\alpha,\theta)}{\partial \theta}
\right|_{\alpha=\pi/2,\theta=0}=
\frac{1}{2}\sum_{k=0}^\infty r^{k}\sum_{p=0}^{[k/2]}\ln^p r \sum_{j=p}^{k-p}Z^{j}
\psi_{k,p}^{(j)}\left(\frac{\pi}{2},0\right).~~~~
\end{equation}
Equating coefficients for the same powers of $r,\ln r$ and $Z$ in both sides of Eq.(\ref{89}), one obtains the relation
\begin{equation}\label{90}
\psi_{k,p}^{(j)}\left(\frac{\pi}{2},0\right)=2\sqrt{2}
\left.\frac{\partial\psi_{k+1,p}^{(j)}(\alpha,\theta)}
{\partial \theta}\right|_{\alpha=\pi/2,\theta=0}
\end{equation}
for the AFC components at the EEC line.


The important features of the limits under consideration have to be reported.
It can be verified that any component/subcomponent $\psi(\alpha,\theta)\equiv \psi_{k,p}^{(j)}(\alpha,\theta)$ can be represented in the form (\ref{30}), where some of the functions $\omega_l^{(k,p;j)}(\alpha)$ can be constants including zero.
For example, using representation (19)\cite{LEZ1} for $\psi_{4,2}(\alpha,\theta)$, one can write down
\begin{equation*}
\omega_l^{(4,2;2)}(\alpha)=\frac{(\pi-2)(5\pi-14)}{180\pi^2}\left[\left(1-\frac{4}{3}\sin^2\alpha\right)\delta_{l0}+\frac{4}{3}\delta_{l2}\right],
\end{equation*}
where $\delta_{lm}$ is the Kronecker delta.
It follows from the form of the rhs of Eq.(\ref{30})
that the partial derivative of the function represented by Eq.(\ref{30}) possesses the property:
\begin{equation}\label{23}
\left.\frac{\partial\psi(\alpha,\theta)}{\partial\theta}\right|_{\alpha=\alpha_0,~\theta=\theta_0}=
\lim_{\theta\rightarrow \theta_0}\frac{d \psi(\alpha_0,\theta)}{d \theta}
.
\end{equation}
It is clear that the equation like (\ref{23}) but for the partial derivatives (not mixed) of the higher orders can be written.
Thereby, one can conclude that the AFC components/subcomponents  possess the property (\ref{23}).
The important remark is: given that the hyperspherical angle $\theta$ is non-negative, relations (\ref{23}) for $\theta_0=0$ reduces to the form
\begin{equation}\label{24}
\left.\frac{\partial\psi(\alpha,\theta)}{\partial\theta}\right|_{\alpha=\alpha_0,~\theta=0}=
\lim_{\theta\rightarrow 0^+}\frac{d \psi(\alpha_0,\theta)}{d \theta}.
\end{equation}
It will be shown that the right-hand side limit in the rhs of Eq.(\ref{24}) is of a special importance for the single series (\ref{30}).

It is well-known that the two-electron wave function (\ref{I1}) and its (at least) first and second
partial derivatives in respect to the interparticle coordinates $r_1,r_2$ and $r_{12}$ must be finite.
Considering the relation (\ref{87}) at the ENC, one obtains
\begin{equation}\label{98}
\left.\frac{\partial r}{\partial r_{12}}\right|_{r_{2}=0}=0,~~
\left.\frac{\partial \alpha}{\partial r_{12}}\right|_{r_{2}=0}=0,~~
\left.\frac{\partial \theta}{\partial r_{12}}\right|_{r_{2}=0}=\infty.
\end{equation}
It follows from Eqs.(\ref{87}) and (\ref{98}) that one should set
\begin{equation}\label{99}
\left.\frac{\partial \Psi}{\partial \theta}\right|_{\alpha=0,\theta=\pi/2}=0
\end{equation}
in order to preserve the finiteness of the partial derivative $\partial\Psi/\partial r_{12}$ at the ENC.
Using the Fock expansion (\ref{I1}), the property (\ref{23}) and the $Z$-power separation (\ref{I11}), one obtains for the AFC components:
\begin{equation}\label{100}
\lim_{\theta\rightarrow\pi/2}\frac{d \psi_{k,p}^{(j)}(0,\theta)}{d \theta}=0.
\end{equation}
Making use of the Eqs.(\ref{24}) and (\ref{100}), one can rewrite the specific conditions (\ref{85}) and (\ref{90}) in the final form
\begin{equation}\label{85a}
\lim_{\theta\rightarrow\pi/2}~\lim_{\alpha\rightarrow0}\frac{\partial}{\partial \alpha}\psi_{k+1,p}^{(j+1)}(\alpha,\theta)=
-\frac{1}{2}\psi_{k,p}^{(j)}\left(0,\frac{\pi}{2}\right),
\end{equation}
\begin{equation}\label{90a}
\lim_{\theta\rightarrow 0^+}\frac{d}{d \theta}\psi_{k+1,p}^{(j)}\left(\frac{\pi}{2},\theta\right)
=\frac{1}{2\sqrt{2}}\psi_{k,p}^{(j)}\left(\frac{\pi}{2},0\right).
\end{equation}
Note that we didn't use $d\psi_{k+1,p}^{(j+1)}(\alpha,\pi/2)/d \alpha$ instead of
$\lim_{\theta\rightarrow \pi/2}\partial\psi_{k+1,p}^{(j+1)}(\alpha,\theta)/\partial \alpha$ in the lhs of
Eq.(\ref{85a}), because, in general case, the limit over angle $\alpha$ must be taken first.

One should emphasize that Eqs.(\ref{85a}) and (\ref{90a}) are derived for the first time,
and represent the specific boundary conditions for the AFC components at the two-particle coalescences.

A number of the AFC components derived in  \cite{LEZ1}, refined and derived in \cite{LEZ2}, and supplemented here are in our disposal.
We have verified the validity of Eqs.(\ref{85a}), (\ref{90a}) on examples of all the presently determined components.
Notice that the presently calculated subcomponents at the two-particle coalescences are presented in Table \ref{T4},
whereas the presently determined components (at the two-particle coalescences) in the right-hand sides
of Eqs.(\ref{85a}), (\ref{90a}) are presented in Tables \ref{T1}-\ref{T3}.
One should underline that in all cases when we could provide verification, the correctness of the boundary
conditions (\ref{85a}), (\ref{90a}) was certified.


Note that there are the components that can be placed to the lhs of Eqs.(\ref{85a}), (\ref{90a}),
while the corresponding AFC components in the rhs, as we could say, do not exist.
However, when we say that they do not exist, we indeed mean the components
$\psi_{k,p}^{(j)}\equiv0$ for $k<0 \vee p>[k/2]\vee j<p\vee j>k-p$.
We have verified that in all those cases Eqs.(\ref{85a}), (\ref{90a}) remain correct.

\begin{figure}
\centering
\mbox{\subfigure{\includegraphics[width=3.3in]{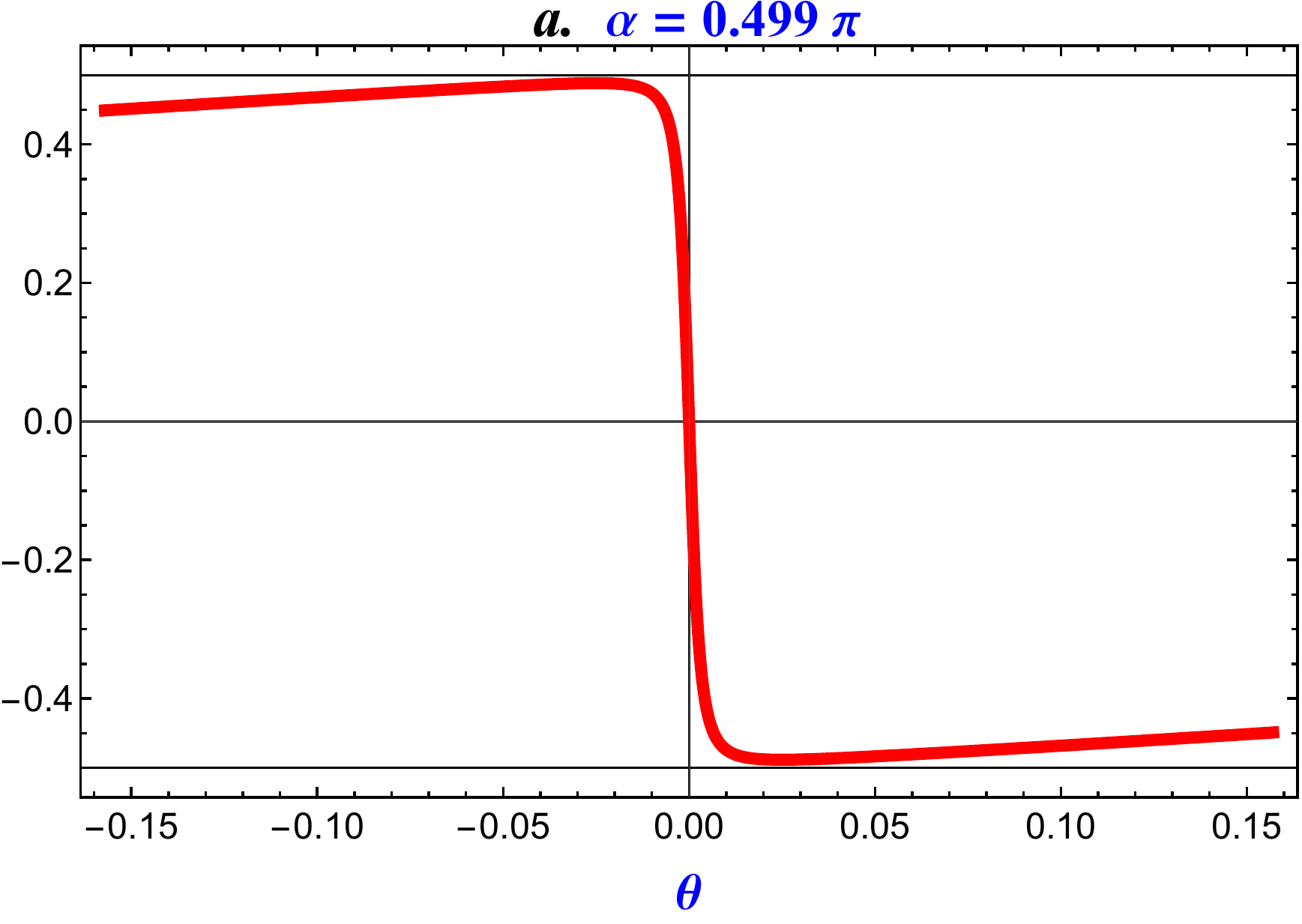}
\quad
\subfigure{\includegraphics[width=3.5in]{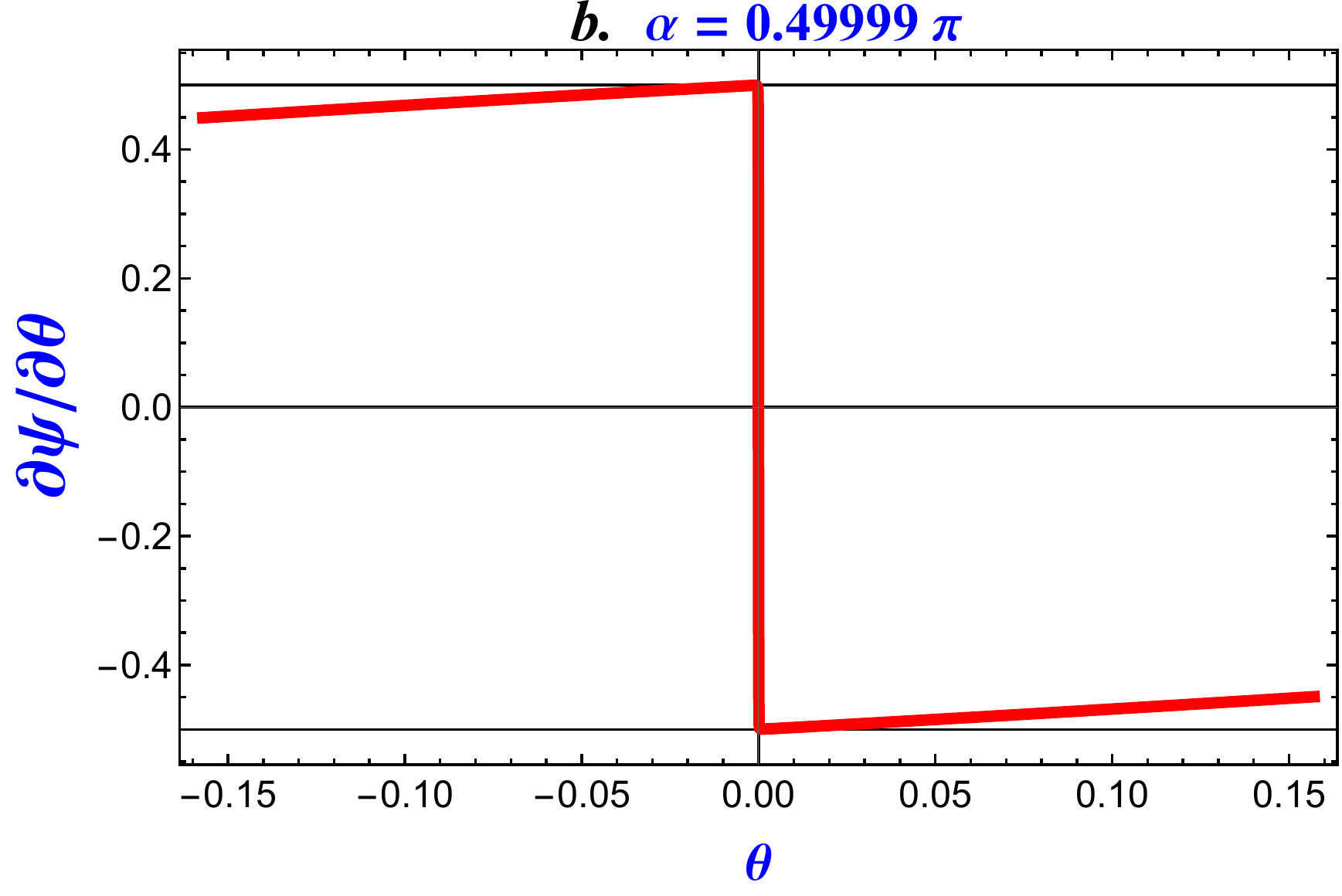} }}}
\caption{Partial derivative $\partial\psi_{2,0}^{(1)}(\alpha,\theta)/\partial\theta$ for different values of $\alpha$ closed to $\pi/2$.}
\label{F1}
\end{figure}

To verify the correctness of the conditions (\ref{85a}), (\ref{90a}), we need to calculate the AFC
components/subcomponents and thier partial derivatives
with respect to the hyperspherical angles $\alpha$ and $\theta$, at the two-particle coalescences.
There is no problem to perform such calculations for the component/subcomponent represented by explicit analytic function
of $\alpha$ and $\theta$, that is when only limited number of the functions $\omega_l(\rho)\equiv\omega_l^{(k,p;j)}(\alpha)$ are different from zero.
However, in case of representation
by infinite single series of the form (\ref{30}), a special consideration is required.
For such cases, it is easy to verify the correctness of the following relations:
\begin{equation}\label{91}
\psi\left(0,\frac{\pi}{2}\right)=\omega_0(0),
\end{equation}
\begin{equation}\label{92}
\psi\left(\frac{\pi}{2},0\right)=\sum_{l=0}^\infty\omega_l(1),
\end{equation}
\begin{equation}\label{93}
\left.\frac{\partial\psi(\alpha,\theta)}{\partial \alpha}\right|_{\alpha=0,\theta=\pi/2}=
\left.\frac{d\omega_0(\rho)}{d\alpha}\right|_{\alpha=0},
\end{equation}
\begin{equation}\label{94}
\left.\frac{\partial\psi(\alpha,\theta)}{\partial \theta}\right|_{\alpha=0,\theta=\pi/2}=0,
\end{equation}
where $\rho$ is defined by Eq.(\ref{31b}).
Note that Eq.(\ref{94}) derived for the single series or HH representations, coincides certainly with the general Eq.(\ref{100}).
The problem is  to calculate (performing term by term differentiation) the derivative in respect to $\theta$
\begin{equation}\label{95}
g(\theta)\equiv\frac{d}{d \theta}\psi\left(\frac{\pi}{2},\theta\right)=
\frac{1}{\sin \theta}\sum_{l=1}^\infty\omega_l\left(\frac{\pi}{2}\right)(l+1)
\left[ P_{l+1}(\cos\theta)-\cos\theta P_l(\cos\theta)\right],
\end{equation}
which, at the first sight, equals zero at the EEC ($\theta=0$).
However, it is not correct in general. Figure \ref{F1} ($a$ and $b$) demonstrates the reason of the possible wrong result.
On this figure, we depicted the plots of the partial derivative $\partial\psi_{2,0}^{(1)}(\alpha,\theta)/\partial\theta$ in
the vicinity of the EEC angle point $\alpha=\pi/2,\theta=0$. To build the plots we used the analytic representation (22)\cite{LEZ1}.
It is seen that the mentioned derivative has a cusp (singularity) at the EEC point of the hyperspherical angular space.
For $\theta\geq0$ we need to calculate the right-hand side limit.

It can be shown that for the components/subcomponents represented by series of the form (\ref{30}) the two-sided limits of
the partial derivative in respect to $\theta$ at the EEC can be calculated in the form
\begin{equation}\label{108a}
\lim_{\theta\rightarrow 0^{\pm}}\frac{d }{d \theta}\psi\left(\frac{\pi}{2},\theta\right)=
\lim_{l\rightarrow\infty}\left\{\mp\omega_l\left(1\right)l^2\right\}.
\end{equation}
Notice that relation (\ref{108a}) remains correct in case of the absence of the cusp for the derivative $g(\theta)$ at
the point $\theta=0$,  that is in case of the function $g(\theta)$  is continues at $\theta=0$, and the left-hand side and the right-hand side
limits are coincident.

Only four subcomponents represented by single series of the form (\ref{30}) were considered in the previous Section and in the
papers \cite{LEZ1,LEZ2}. These are $\chi_{20},\psi_{4,1}^{(2d)},\psi_{3,0}^{(2e)}$ and $\psi_{3,0}^{(2c)}$.
Using the representation (\ref{95}) with the increasing (but finite) upper limits, we can build the sequence of plots
for the derivatives $g(\theta)$ to ensure that the corresponding limits are
\begin{equation}\label{96}
\lim_{l\rightarrow\infty}\left\{-\sigma_l\left(1\right)l^2\right\}=-\frac{1}{6},~~~
\lim_{l\rightarrow\infty}\left\{-\tau_l\left(1\right)l^2\right\}=\frac{\pi-2}{12\pi},~~~
\lim_{l\rightarrow\infty}\left\{-\phi_l\left(1\right)l^2\right\}=0,
\end{equation}
where $\sigma_l(1),\tau_l(1)$ and $\phi_l(1)$ are defined by Eqs.(\ref{37}),(\ref{56}) and (\ref{77}), respectively.
It is easy to verify that the results (\ref{96}) confirm the validity of the conditions
\begin{equation}\label{108}
\lim_{\theta\rightarrow 0^{+}}\frac{d}{d\theta}\psi_{2,0}^{(1)}\left(\frac{\pi}{2},\theta\right)=
\frac{1}{2\sqrt{2}}\psi_{1,0}^{(1)}\left(\frac{\pi}{2},0\right)=-\frac{1}{2},
\end{equation}
\begin{equation}\label{109}
\lim_{\theta\rightarrow 0^{+}}\frac{d}{d\theta}\psi_{4,1}^{(2)}\left(\frac{\pi}{2},\theta\right)=
\frac{1}{2\sqrt{2}}\psi_{3,1}^{(2)}\left(\frac{\pi}{2},0\right)=\frac{\pi-2}{12\pi},
\end{equation}
corresponding to the general formula (\ref{90a}) for the EEC.
The rhs of Eq.(\ref{108}) can be also obtained by direct using the analytic representation (22)\cite{LEZ1}
for the function $\psi_{2,0}^{(1)}$ in the lhs.
Moreover, using Eq.(\ref{93}) and the results of Appendix \ref{SA},
one easily obtains the boundary condition
\begin{equation}\label{109a}
\lim_{\theta\rightarrow \pi/2}\lim_{\alpha\rightarrow 0}\frac{d\psi_{3,0}^{(2)}\left(\alpha,\theta\right)}{d\alpha}=
-\frac{1}{2}\psi_{2,0}^{(1)}\left(0,\frac{\pi}{2}\right)=\frac{3-\ln 2}{12}
\end{equation}
at the ENC.

In analogy to Eqs.(\ref{87}) and (\ref{98})-(\ref{99}) it is easy to write down the chain rule relation for the second
derivative $\partial^2\Psi/\partial r_{12}^2$.
This relation, taken at the ENC line, enables us to obtain the condition
\begin{equation}\label{101}
\lim_{\theta\rightarrow\pi/2}\frac{d^2 \psi_{k,p}^{(j)}(0,\theta)}{d \theta^2}=0.
\end{equation}
The relation
\begin{equation}\label{102}
2\left.\frac{\partial\chi(r_1,r_2,r_{12})}{\partial r_2}\right|_{r_1=r_2=R}=
\frac{\partial \chi\left(R,R,r_{12}\right)}{\partial R}
\end{equation}
was obtained previously (see Eq.(40)\cite{LEZ0}).
Considering the relation (\ref{82}) at the EEC, one obtains
\begin{equation}\label{104}
\left.\frac{\partial r}{\partial r_2}\right|_{r_{12}=0}=\frac{1}{\sqrt{2}},~~
\left.\frac{\partial \alpha}{\partial r_2}\right|_{r_{12}=0}=\frac{\sqrt{2}}{r},~~
\left.\frac{\partial \theta}{\partial r_{2}}\right|_{r_{12}=0}=0.
\end{equation}
Substitution of Eq.(\ref{82}) taken at the EEC, and Eq.(\ref{104}) into Eq.(\ref{102}) with $r_{12}=0 \wedge R=r/\sqrt{2}$ yields
\begin{equation}\label{105}
\left.\left(\frac{\partial \Psi}{\partial r}+\frac{2}{r}\frac{\partial \Psi}{\partial \alpha}\right)
\right|_{\alpha=\pi/2,\theta=0}=\frac{d}{d r}\Psi\left(r,\frac{\pi}{2},0\right),
\end{equation}
whence one obtains the condition
\begin{equation}\label{106}
\left.\frac{\partial \Psi}{\partial \alpha}\right|_{\alpha=\pi/2,\theta=0}=0.
\end{equation}
Substitution of the Fock expansion (\ref{I1}) and the relation (\ref{I11}) into Eq.(\ref{106}) yields for  derivative of the AFC
components at the EEC:
\begin{equation}\label{107}
\lim_{\theta\rightarrow 0}\lim_{\alpha\rightarrow \pi/2}\frac{\partial }{\partial \alpha}\psi_{k,p}^{(j)}(\alpha,\theta)=0.
\end{equation}
The correctness of the relations (\ref{100}), (\ref{101}) and (\ref{107}) was verified on examples of the presently determined AFC components.

\section{Conclusions}\label{S4}

The exact values of all the presently determined AFC components at the two-particle coalescences were derived and presented in Tables \ref{T1}-\ref{T4}.
The corresponding results for the edge components $\psi_{k,0}^{(0)}$ and $\psi_{k,0}^{(k)}$ with $1\leq k\leq8$ were displayed in Tables \ref{T2}
and \ref{T3}, respectively.
The infinite series summation by \emph{Mathematica} and the use of the \emph{Mathematica} operator \textbf{FindSequenceFunction} enabled us to obtain the exact results at the EEC even for the components/subcomponents represented by infinite single series of the form (\ref{30}).
As a side effect, it was found the exact "pure" analytical representation for subcomponent $\chi_{20}(\alpha,\theta)$ obtained by use of the exact admixture
coefficient defined by Eq.(\ref{43}).
The Green function approach has proven to be useful as the additional method of calculation of the AFC components at the two-particle coalescences.

The boundary conditions (\ref{85a}) and (\ref{90a}) for the AFC components, expressed through the hyperspherical angles $\alpha$ and $\theta$, were derived as the result of application of the Kato cusp conditions for the two-electron ($S$-state) wave function expressed
in the interparticle distances $r_1,r_2$ and $r_{12}$.
The additional boundary conditions that have nothing to do with the Kato cusp conditions, were obtained, as well.
The correctness of the obtained boundary conditions was verified on examples of all the presently determined AFC components.

\section{Acknowledgment}

The author acknowledges Prof. Nir Barnea for useful discussions.
The author is grateful to Prof. Paul Abbott for providing his thesis.
This work was supported by the PAZY Foundation.

\appendix

\section{}\label{SA}

It was shown in \cite{LEZ1} that in order to derive the physical solution of the IFRR (\ref{205}) in the form of the single
 series (\ref{209}), one should, first of all, to represent the rhs (\ref{206}) in the similar form
\begin{equation}\label{A1}
h_{3,0}^{(2d)}=\sum_{l=0}^\infty P_l(\cos \theta)\left(\sin \alpha\right)^lh_l(\rho).
\end{equation}
Then, the basis functions $g_l(\rho)$ of the single-series representation (\ref{209}) can be found as the physical solution of the equation
\begin{equation}\label{A2}
\left(1+\rho^2\right)^2g_l''(\rho)+2\rho^{-1}\left[1+\rho^2+l(1-\rho^4)\right]g_l'(\rho)+(3-2l)(2l+7)g_l(\rho)=-h_l(\rho).
\end{equation}
The individual solutions $u_{3l}(\rho)$ and $v_{3l}(\rho)$ of the homogeneous equation associated with Eq.(\ref{A2})
 are represented by Eqs.(\ref{12}) and (\ref{65}), respectively. Using the method of variation of parameters, the particular solution
  of the inhomogeneous equation (\ref{A2}) can be found in the form \cite{LEZ1}:
\begin{equation}\label{A3}
g_l^{(p)}(\rho)=\frac{1}{2l+1}\left[u_{3l}(\rho)\int\frac{v_{3l}(\rho)h_l(\rho)\rho^{2l+2}}{(\rho^2+1)^{2l+3}}d\rho-
v_{3l}(\rho)\int\frac{u_{3l}(\rho)h_l(\rho)\rho^{2l+2}}{(\rho^2+1)^{2l+3}}d\rho
\right].
\end{equation}
We shall find only the function $g_0(\rho)$ used in the main text of this paper.
Making use of definition (\ref{206}), one obtains (see also Eq.(63) \cite{LEZ1}) the expression for the rhs
\begin{equation}\label{A4}
h_0(\rho)=\frac{1+\rho}{3\rho\sqrt{1+\rho^2}}\left[
(1-\rho^2)\ln\left(\frac{2}{1+\rho^2}\right)-\frac{(\rho^4-6\rho^2+1)\arctan(\rho)}{2\rho}
-\frac{1}{2}(3\rho^2+2\rho+1)
\right],
\end{equation}
which is correct for $0\leq\rho\leq 1$.
For $\rho >1$ one should replace $\rho$ by $1/\rho$ in the rhs of Eq.(\ref{A4}).
The formula (\ref{A3}) gives the corresponding particular solution in the following rather complicated form
\begin{eqnarray}\label{A5}
g_0^{(p_1)}(\rho)=\frac{(\rho^2+1)^{-3/2}}{360\rho}\left\{-6\ic\left[(5\rho^4-10\rho^2+1)Li_2\left(-e^{2\ic \arctan \rho}\right)+
\rho(\rho^4-10\rho^2+5)Li_2\left(e^{2\ic \arctan \rho}\right)\right]
\right.
~\nonumber~~~\\
+20\rho(3\rho^2+\rho-2)\ln\left(\frac{2}{1+\rho^2}\right)+8\ln(1+\rho^2)+\arctan(\rho)\left[
12(5\rho^4-10\rho^2+1)\ln\left(\frac{2\ic}{\rho+\ic}\right)+
\right.
~~~~\nonumber\\
\left.
12\rho(\rho^4-10\rho^2+5)\ln\left(\frac{2\rho}{\rho+\ic}\right)-\rho(7\rho^4-77\rho^3-46\rho^2+114\rho+3)
+6(1-\ic)(\rho-\ic)^5\arctan(\rho)+9\right]
~\nonumber~\\
\left.
-\rho(7\rho^3-11\rho^2-81\rho-31)-8(1+\ln 2)\right\},~~~~~~~~~~~~~~~~~~~~~~
\end{eqnarray}
where $Li_2(z)$ is the dilogarithm function and $\ic=\sqrt{-1}$. The problem is that the particular solution (\ref{A5})
 is singular and complex in the vicinity of the point $\rho=0$, in particular
\begin{eqnarray}\label{A6}
g_0^{(p_1)}(\rho)\underset{\rho\rightarrow 0}{=}
\frac{1}{720\rho}\left[\ic\pi^2-16(1+\ln 2)\right]-\frac{1}{72}\left[\ic \pi^2+8(1-\ln 2)\right]-
~\nonumber~~~~~~~~~~~~~~~~\\
\frac{\rho}{1440}(23\ic\pi^2-608-128\ln 2)+O(\rho^2).~~~~~~~~~~~
\end{eqnarray}
It is more convenient to use the real and finite particular solution that can be obtained by transformation
\begin{equation}\label{A7}
g_0^{(p_2)}(\rho)=g_0^{(p_1)}(\rho)+\frac{\ic\pi^2}{72}v_{30}(\rho)+\frac{16(1+\ln 2)-\ic\pi^2}{720}u_{30}(\rho).
\end{equation}
The power series expansion for the latter particular solution reads
\begin{equation}\label{A8}
g_0^{(p_2)}(\rho)\underset{\rho\rightarrow 0}{=}
\frac{1}{9}(1-\ln 2)+\frac{\rho}{6}(1-\ln 2)+\frac{\rho^2}{18}(6\ln 2-5)+O(\rho^3).
\end{equation}
The physical solution of Eq.(\ref{A2}) with $l=0$ can be calculated in the form \cite{LEZ1}
\begin{equation}\label{A9}
g_0(\rho)=g_0^{(p_2)}(\rho)+c_0 v_{30}(\rho).
\end{equation}
In order to find the coefficient $c_0$ we apply the coupling equation (61)\cite{LEZ1} which for this case becomes
\begin{equation}\label{A10}
\mathcal{F}_{0,0}=\frac{32}{\pi}\int_0^1 g_0(\rho)\frac{\rho^2}{(\rho^2+1)^3}d\rho,
\end{equation}
where $\mathcal{F}_{n,l}$ represents the unnormalized HH expansion coefficient for subcomponent
\begin{equation}\label{A11}
\psi_{3,0}^{(2d)}(\alpha,\theta)=\sum_{l=0}^\infty\mathcal{F}_{n,l}Y_{nl}(\alpha,\theta).
\end{equation}
It was shown (see Appendix D \cite{LEZ1}) that for the AFC $\psi_{k,p}^{(j)}$ with $k=3$, the following relation is valid
\begin{equation}\label{A12}
\mathcal{F}_{n,l}=\frac{\mathcal{H}_{n,l}}{(n-3)(n+7)},
\end{equation}
where
\begin{equation}\label{A13}
\mathcal{H}_{n,l}=N_{nl}^2\int h_{3,0}^{(2d)}(\alpha,\theta)Y_{nl}(\alpha,\theta)d \Omega
\end{equation}
represents the coefficient of the HH expansion of the rhs $h_{3,0}^{(2d)}$ defined by Eqs.(\ref{206}) and (\ref{A1}).
The normalization coefficient $N_{nl}$ and the volume element $d\Omega$ are defined in Ref.\cite{LEZ1}.
Thus, inserting Eq.(\ref{A13}) into the rhs of Eq.(\ref{A12}) with $n=l=0$, one obtains
\begin{equation}\label{A14}
\mathcal{F}_{0,0}=-\frac{2N_{00}^2\pi^2}{21}\int_0^\pi h_0(\rho)\sin^2\alpha~d\alpha=
-\frac{32\left[2\sqrt{2}+\ln(6-4\sqrt{2})-5\right]}{945\pi}.
\end{equation}
Equating the right-hand sides of Eqs.(\ref{A10}) and (\ref{A14}) and using Eq.(\ref{A9}), one finds the required coefficient in the form
\begin{eqnarray}\label{A15}
c_0=-\left(\int_0^1 v_{30}(\rho)\frac{\rho^2}{(\rho^2+1)^3}d\rho\right)^{-1}\left[
\frac{2\sqrt{2}+\ln(6-4\sqrt{2})-5}{945}+\int_0^1g_0^{(p_2)}(\rho)\frac{\rho^2}{(\rho^2+1)^3}d\rho
\right]
~\nonumber~~~\\
\simeq-0.0316978107237.~~~~~~~~~~~~~~~~~~~~~~~~~~~~~~~~~~~~~~~~~~~~~
\end{eqnarray}
It is seen from the complicated representation (\ref{A5}) that it is very difficult to derive the exact value of $c_0$ using Eq.(\ref{A15}).
However, there is no problem to calculate its numerical value.
Thus, substitution of the results of this Appendix into Eq.(\ref{210}) yields the following numerical value for the component
$\psi_{3,0}^{(2d)}$ at the ENC:
\begin{equation}\label{A16}
\psi_{3,0}^{(2d)}\left(0,\frac{\pi}{2}\right)=g_0^{(p_2)}(0)+c_0v_{30}(0)=\frac{1-\ln 2}{9}+c_0=0.002396946991882.
\end{equation}
It is easy to verify that the numerical value of the exact representation (\ref{211}) obtained in Sec.\ref{S2} coincides
with the numerical result (\ref{A16}), as expected. This enables us among other things to establish the exact value of the coefficient
\begin{equation}\label{A17}
c_0=\frac{\pi(16-3\pi)-48G-8+32 \ln 2}{288}.
\end{equation}

\newpage

\begin{table}
\begin{center}
\caption{The components of the AFC at the two-particle coalescences.  }
\begin{tabular}[t]{|c|c|c|c|c|}
\hline
$k$ & $p$ & $j$ & $\psi{}_{k,p}^{(j)}(0,\frac{\pi}{2})$ & $\psi{}_{k,p}^{(j)}(\frac{\pi}{2},0)$\tabularnewline
\hline
0 & 0 & 0 & $1$ & 1\tabularnewline
\hline
2 & 0 & 1 & $\frac{\ln2-3}{6}$ & $\frac{48G-62+\pi(5+12\ln2)}{72\pi}$\tabularnewline
\hline
2 & 1 & 1 & $0$ & $\frac{2-\pi}{3\pi}$\tabularnewline
\hline
3 & 1 & 1 & $\frac{2-\pi}{36\pi}$ & $0$\tabularnewline
\hline
3 & 1 & 2 & $0$ & $\frac{\pi-2}{3\pi\sqrt{2}}$\tabularnewline
\hline
4 & 1 & 1 & $\frac{2-\pi}{576\pi}$ & $\frac{(\pi-2)(32E-15)}{960\pi}$\tabularnewline
\hline
4 & 1 & 2 & $\frac{(\pi-2)[424-600G+25\pi(12\ln2-7)]}{5400\pi^{2}}$ & $\frac{(\pi-2)[11536-8400G+\pi(2205\pi-17294+2100\ln2)]}{75600\pi^{2}}$\tabularnewline
\hline
4 & 1 & 3 & $0$ & $\frac{3(2-\pi)}{40\pi}$\tabularnewline
\hline
4 & 2 & 2 & $\frac{(\pi-2)(5\pi-14)}{180\pi^{2}}$ & $\frac{(\pi-2)(5\pi-14)}{180\pi^{2}}$\tabularnewline
\hline
\end{tabular}
\label{T1}
\end{center}
\end{table}

\begin{table}
\begin{center}
\caption{The lower edge components $\psi_{k,0}^{(0)}$ of the the logarithmless AFC at the two-particle coalescences.  }
\begin{tabular}[t]{|c|c|c|}
\hline
$k$ & $\psi{}_{k,0}^{(0)}(0,\frac{\pi}{2})$ & $\psi{}_{k,0}^{(0)}(\frac{\pi}{2},0)$ \tabularnewline
\hline
1 & $\frac{1}{2}$ & $0$ \tabularnewline
\hline
2 & $\frac{1}{12}(1-2E)$ &  $\frac{1}{12}(1-2E)$  \tabularnewline
\hline
3 & $\frac{1-5E}{72}$ & $0$  \tabularnewline
\hline
4 & $\frac{12E^2-11E+1}{1152}$ &$\frac{20E^2-21E+7}{1920}$  \tabularnewline
\hline
5 & $\frac{160E^2-47E-6}{43200}$ &$0$  \tabularnewline
\hline
6 & $\frac{-1260E^3+1595E^2-189E-62}{3628800}$ &$\frac{-10080E^3+16460E^2-11361E+3007}{29030400}$  \tabularnewline
\hline
7 & $\frac{-786240E^3+261908E^2+100023E-27233}{7315660800}$ &$0$ \tabularnewline
\hline
8 & $\frac{16934400E^4-26481840E^3+3072680E^2+2788023E-325061}{2341011456000}$ &$\frac{30481920E^4-68266800E^3+72613544E^2-39544113E+8871475}{4213820620800}$\tabularnewline
\hline
\end{tabular}
\label{T2}
\end{center}
\end{table}

\begin{table}
\begin{center}
\caption{The higher edge components $\psi_{k,0}^{(k)}$ of the the logarithmless AFC at the two-particle coalescences.  }
\begin{tabular}[t]{|c|c|c|}
\hline
$k$ & $\psi{}_{k,0}^{(k)}(0,\frac{\pi}{2})$ & $\psi{}_{k,0}^{(k)}(\frac{\pi}{2},0)$ \tabularnewline
\hline
1 & $-1$ & $-\sqrt{2}$ \tabularnewline
\hline
2 & $\frac{1}{3}$ &  $\frac{5}{6}$  \tabularnewline
\hline
3 & $-\frac{1}{18}$ & $-\frac{7\sqrt{2}}{36}$  \tabularnewline
\hline
4 & $\frac{7}{288}+\frac{2}{45\pi}$ &$\frac{5}{96}-\frac{2}{135\pi}$  \tabularnewline
\hline
5 & $-\frac{43}{4320}-\frac{46}{2025\pi}$ &$\frac{1}{\sqrt{2}}\left(-\frac{1}{180}+\frac{32}{2025\pi}\right)$  \tabularnewline
\hline
6 & $\frac{89}{45360}+\frac{28}{6075\pi}$ &$-\frac{83}{120960}-\frac{47}{12150\pi}$  \tabularnewline
\hline
7 & $-\frac{139}{635040}-\frac{22}{42525\pi}$ &$\frac{1}{\sqrt{2}}\left(\frac{377}{1016064}+\frac{13}{11340\pi}\right)$ \tabularnewline
\hline
8 & $\frac{25507}{390168576}+\frac{2524019}{9601804800\pi}+\frac{412}{1607445\pi^2}$
  & $-\frac{21871}{650280960}-\frac{679193}{9601804800\pi}+\frac{412}{8037225\pi^2}$\tabularnewline
\hline
\end{tabular}
\label{T3}
\end{center}
\end{table}

\begin{table}
\begin{center}
\caption{The subcomponents of the AFC with $k>2$ at the two-particle coalescences.  }
\begin{tabular}{|c|c|c|c|c|}
\hline
$k$ & $p$ & $j$ & $\psi{}_{k,p}^{(j)}(0,\frac{\pi}{2})$ & $\psi{}_{k,p}^{(j)}(\frac{\pi}{2},0)$\tabularnewline
\hline
\hline
3 & 0 & 1$a$ & $\frac{5(\pi-2)}{72\pi}$ & $0$\tabularnewline
\hline
3 & 0 & 1$b$ & $\frac{4E-1}{36}$ & $\frac{5E-2}{18\sqrt{2}}$\tabularnewline
\hline
3 & 0 & 1$c$ & $\frac{5(2-\pi)(20+3\pi)}{1728\pi}$ & $\frac{25(\pi-2)}{288\sqrt{2}}$\tabularnewline
\hline
\hline
3 & 0 & 2$a$ & $\frac{1}{9}$ & $0$\tabularnewline
\hline
3 & 0 & 2$b$ & $0$ & $\frac{(2-\pi)(36+5\pi)}{144\pi\sqrt{2}}$\tabularnewline
\hline
3 & 0 & 2$c$ & $\frac{1}{18}(2-\pi-2\ln2)$ & $\frac{\sqrt{2}}{27}$\tabularnewline
\hline
3 & 0 & 2$d$ & $\frac{24-48G+\pi(16-3\pi)}{288}$ & $\ldots$\tabularnewline
\hline
3 & 0 & 2$e$ & $\frac{1}{8}$ & $\frac{5-6 G}{12\sqrt{2}}$\tabularnewline
\hline
\hline
3 & 0 & $2$ & $\frac{124-48G-3\pi^2-32\ln 2}{288}$ & $\ldots$\tabularnewline
\hline
\hline
4 & 1 & 2$b$ & $\frac{(\pi-2)(5\pi-14)}{1080\pi^{2}}$ & $\frac{(\pi-2)(5\pi-14)}{3240\pi^{2}}$\tabularnewline
\hline
4 & 1 & 2$c$ & $0$ & -$\frac{37(\pi-2)(5\pi-14)}{8100\pi^{2}}$\tabularnewline
\hline
4 & 1 & 2$d$ & $\frac{(\pi-2)[247-300G+50\pi(3\ln2-2)]}{2700\pi^{2}}$ & $\frac{(\pi-2)[7028-8400G+3\pi(735\pi-5228+700\ln2)]}{75600\pi^{2}}$\tabularnewline
\hline
\hline
4 & 1 & $2$ & $\frac{(\pi-2)[424-600G+25\pi(12\ln2-7)]}{5400\pi^{2}}$ & $\frac{(\pi-2)[11536-8400G+\pi(2205\pi-17294+2100\ln2)]}{75600\pi^{2}}$\tabularnewline
\hline
\end{tabular}
\label{T4}
\end{center}
\end{table}

\end{document}